\documentclass[conference]{IEEEtran}
\IEEEoverridecommandlockouts
\usepackage{cite}
\usepackage{amsmath,amssymb,amsfonts}
\usepackage{booktabs}
\usepackage{multirow}
\usepackage[linesnumbered,ruled,vlined]{algorithm2e}
\usepackage{graphicx}
\usepackage{subcaption}
\usepackage{textcomp}
\usepackage{xcolor}
\usepackage{multicol}
\usepackage{balance}
\def\BibTeX{{\rm B\kern-.05em{\sc i\kern-.025em b}\kern-.08em
    T\kern-.1667em\lower.7ex\hbox{E}\kern-.125emX}}
\begin{document}

\title{Decision-Aware Semantic State Synchronization in Compute-First Networking\\
{}
}

\author{
Jianpeng Qi,
Chao Liu,
Chengrui Wang,
Rui Wang,
Junyu Dong,
Yanwei Yu%
\thanks{
Jianpeng Qi, Chao Liu, Junyu Dong, and Yanwei Yu are with Ocean University of China, Qingdao, China.
}
\thanks{
Chengrui Wang is with Gosci Technology Group, Qingdao, China.
}
\thanks{
Rui Wang is with University of Science and Technology Beijing, Beijing, China.
}
}

\maketitle

\begin{abstract}
In Compute-First Networking (CFN), an Access Point (AP) makes task offloading decisions based on resource state information reported by a Service Node (SN). A fundamental challenge arises from the trade-off between update overhead and decision accuracy: Frequent state updates consume limited network resources, while infrequent updates lead to stale state views and degraded task performance, especially under high system load. Existing approaches based on periodic updates or Age of Information (AoI) mainly focus on temporal freshness and often overlook whether a state change is actually relevant to offloading decisions.
This paper proposes SenseCFN, a decision-aware state synchronization framework for CFN. Instead of synchronizing raw resource states, SenseCFN focuses on identifying state changes that are likely to alter offloading decisions. To this end, we introduce a lightweight semantic state representation that captures decision-relevant system characteristics, along with a Semantic Deviation Index (SDI) to quantify the impact of state shifts on decision outcomes. Based on SDI, the SN triggers updates only when significant decision-impacting changes are detected. Meanwhile, the AP performs offloading decisions using cached semantic states with explicit awareness of potential staleness. The update and offloading policies are jointly optimized using a centralized training with distributed execution (CTDE) approach.
Simulation results show that SenseCFN maintains a task success rate of up to 99.6\% in saturation-prone scenarios, outperforming baseline methods by more than 25\%, while reducing status update frequency by approximately 70\% to 96\%. These results indicate that decision-aware state synchronization provides an effective and practical alternative to purely time-based update strategies in CFN.
\end{abstract}


\section{Introduction}
With the emergence of ``Compute-First Networking'' (CFN), migrating computing capabilities from centralized clouds to the network edge has become a fundamental paradigm for supporting latency-critical and computation-intensive applications, such as the Metaverse and Extended Reality (XR). In a typical end--edge collaborative scenario, the Access Point (AP, End) acts as a distributed scheduling outpost, which must make task offloading decisions based on resource status periodically reported by the Service Node (SN, Edge) \cite{RN2762,RN2650,JI2025110903}. The correctness of these decisions critically depends on the consistency between the AP’s cached state and the actual execution capability at the SN.

\begin{figure}[htb]
	\centering
	\includegraphics[width=0.9\linewidth]{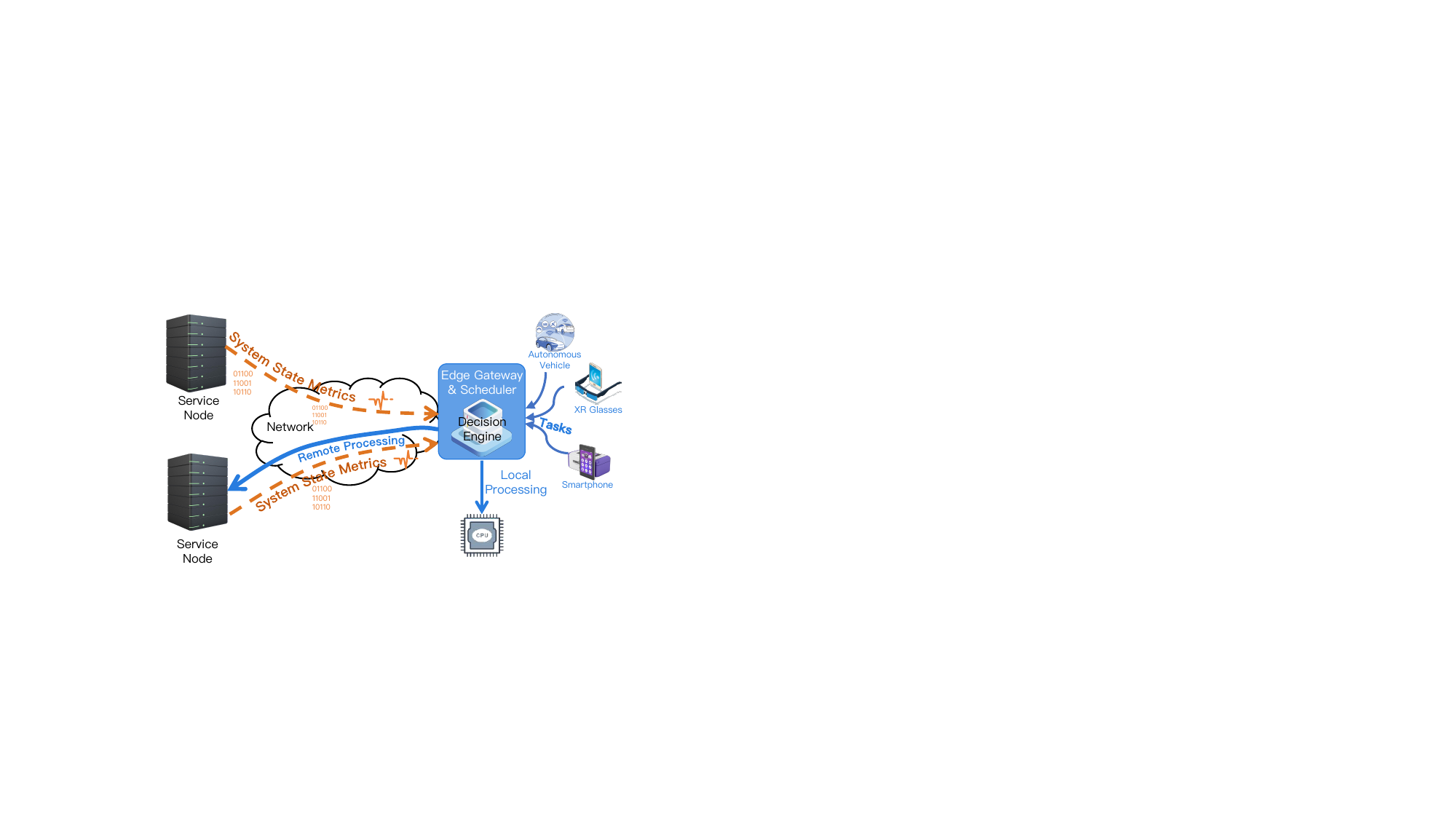}
	\caption{The closed-loop system in CFN.}
	\label{fig:intro-scenario}
\end{figure}

As shown in Fig. \ref{fig:intro-scenario}, the end--edge collaboration in CFN operates through two tightly coupled logical flows: The \textit{Information Flow} for status updating and the \textit{Task Flow} for computation offloading. At the SN, the system continuously monitors dynamic computing resource states, such as CPU core availability and task queue backlog, and selectively transmits this information to the AP via the uplink. At the AP, the decision engine relies on locally cached state information to determine whether an arriving task should be executed locally or offloaded to the SN through the downlink. These two processes form a closed feedback loop, where the effectiveness of status synchronization directly affects the feasibility and timeliness of task execution.

A fundamental challenge in this closed-loop system arises from the tension between decision accuracy and communication overhead. Frequent status updates consume scarce uplink bandwidth and may interfere with service traffic \cite{RN2748}, whereas infrequent updates lead to stale or inconsistent state views at the AP, causing uninformed offloading decisions, increased resource contention, and eventual task deadline violations \cite{huang2025utility}. This trade-off is intrinsic to CFN systems and cannot be resolved solely by increasing update frequency.

The difficulty is further exacerbated by the multidimensional and stochastic nature of edge resource states. Service capability is jointly influenced by heterogeneous factors, including the number of idle cores, queue lengths, and Head-of-Line (HOL) waiting times. These raw physical measurements are coupled in a nonlinear manner, and their numerical values do not translate directly into execution feasibility. For example, a temporary increase in queue length does not necessarily imply excessive latency if sufficient idle cores are available, while a short queue may still conceal severe congestion risks due to HOL blocking.

This ambiguity becomes particularly critical when the system operates near saturation. Under light load, decision errors can often be absorbed by surplus resources. However, in high-load regimes, small fluctuations in resource states may trigger abrupt changes in execution feasibility. Minor numerical variations that appear insignificant in isolation can induce opposite offloading decisions, making distributed decision-making highly sensitive to state inconsistency.

Existing status update mechanisms predominantly address this problem from a temporal freshness perspective. Periodic updates and Age of Information (AoI)-based approaches aim to bound staleness or numerical estimation errors. However, these methods implicitly assume a monotonic relationship between state freshness and decision quality, which does not hold in CFN systems. They fail to distinguish between state changes that are decision-neutral and those that fundamentally alter offloading feasibility \cite{li2024goal}. Consequently, bandwidth resources are often wasted on redundant updates, while critical structural changes in service capability are detected too late.

Motivated by this observation, we revisit status synchronization in CFN from a \emph{decision consistency} perspective. Rather than enforcing uniform freshness, we focus on identifying and transmitting only those state changes that are likely to alter downstream offloading decisions. Under this principle, a status update is necessary only when the discrepancy between the actual and cached states is large enough to induce a decision reversal at the AP.

Based on this insight, we propose \textit{SenseCFN}, a task-oriented end--edge collaborative framework that enables decision-impact-aware state synchronization. In this context, ``semantics'' refers not to content reconstruction fidelity, but to \emph{control semantics}, namely the minimal set of latent state features that structurally affect offloading decisions. A state change is considered significant only if it modifies the decision outcome, rather than merely fluctuating numerically.

SenseCFN operationalizes this principle through two coordinated mechanisms. First, a lightweight semantic encoder abstracts high-dimensional and noisy resource states into compact representations that capture their decision-relevant characteristics. Second, a collaborative dual-decision mechanism jointly governs status updates at the SN and offloading actions at the AP. Updates are triggered only when structurally significant deviations are detected, while the AP performs robust decisions based on cached representations with explicit awareness of potential staleness. These mechanisms are co-optimized using centralized training with distributed execution (CTDE), ensuring low inference overhead and stable online operation.
%
The main contributions are summarized as follows:
\begin{itemize}
\item We propose SenseCFN, a task-oriented end--edge collaborative framework that addresses status synchronization from a decision consistency perspective. By coupling update and offloading decisions, SenseCFN mitigates performance degradation caused by state--decision mismatches in high-load regimes.

\item We design a lightweight decision-aware architecture that integrates semantic state abstraction with a collaborative dual-decision mechanism. By introducing a Semantic Deviation Index (SDI), the system selectively triggers updates only upon structurally significant state shifts, reserving communication resources for high-impact information.

\item Through extensive simulations, we demonstrate that SenseCFN achieves a favorable balance between task success rate and communication overhead, particularly near system saturation, validating the effectiveness of decision-impact-aware state synchronization in resource-constrained edge networks.
\end{itemize}

\section{Related Work}
Resource orchestration in CFN has been extensively studied from the perspectives of communication efficiency, decision accuracy, and system stability. Existing research can be broadly categorized according to how status information is represented, transmitted, and exploited for distributed decision-making. In this work, we organize the related literature along three system-relevant dimensions: (A) State abstraction and information reduction for edge intelligence, (B) decision-aware status update mechanisms that regulate when and what to synchronize, and (C) collaborative offloading strategies that jointly consider communication and computation constraints.

\subsection{Semantic Communication Paradigms for Edge Intelligence}
As traditional ``bit transmission'' models encounter severe bottlenecks under the demands of massive connectivity and wide-area coverage, the research paradigm is actively shifting toward ``semantic and knowledge transmission'' to eliminate redundant information~\cite{meng2024semantics,luo2025informationfreshnesssemanticsinformation,10704713}. Fundamental frameworks like ``Less data, more knowledge''~\cite{yang2022semantic,yang2022semantic-intelli,chaccour2024less,uysal2022semantic} have laid the theoretical foundation for edge-native semantic networks, advocating for the extraction of task-relevant features over raw data streams. Specific lightweight implementations include deep learning-based model pruning for text (L-DeepSC)~\cite{xie2020lite}, adaptive compression for latency-sensitive inference~\cite{liu2023adaptable}, semantic entropy for Quality of Experience (QoE) modeling~\cite{yan2022qoe}, and graph-based coding for efficient image sensing~\cite{kang2023personalized}. Additionally, semantic caching has been explored to further reduce transmission overhead by reusing previously received semantic features~\cite{ZHANG2025111531,10272514}.
\textit{However}, these works predominantly focus on the reconstruction fidelity or task accuracy of multimedia content (e.g., text, audio, images). They largely overlook the unique semantic attributes of resource states (e.g., queue lengths, congestion levels). Unlike multimedia content, the semantic value of resource states lies not in their accurate reconstruction at the receiver, but in their decision impact on downstream control flows. 
This distinction indicates that, unlike content transmission, state representation in CFN must be evaluated by its impact on downstream decision consistency rather than reconstruction accuracy, calling for control-oriented abstractions that align state synchronization with distributed decision outcomes.

\subsection{Decision-Aware Status Update Mechanisms}
Beyond temporal freshness metrics like AoI, research has shifted toward quantifying the ``urgency'' or ``value'' of updates~\cite{11142597}. Advanced metrics such as Age of Incorrect Information (AoII)~\cite{9838737} and goal-oriented sampling strategies~\cite{pappas2021goal,agheli2022semantics} have been proposed to trigger updates only when discrepancies significantly affect receiver inference. Recent advances integrate freshness with utility functions, yielding closed-loop metrics like Goal-oriented Tensor (GoT)~\cite{li2024goal, 10579545}, Age of Semantic Importance (AoSI)~\cite{10570932}, Pareto analysis on communication value~\cite{luo2025value}, and Utility Loss of Information (UoI/SUL)~\cite{huang2025utility, huang2024semantic, xu2024energy}.
\textit{Crucially}, most of these metrics implicitly assume a linear or monotonic relationship between estimation error (e.g., Mean Squared Error) and value loss. This fails to capture the nonlinear ``threshold effect'' inherent in offloading decisions, where minute state fluctuations in critical saturation zones (e.g., near server capacity) can trigger binary decision flips (e.g., from remote processing to local), while massive changes in non-critical zones may have zero impact on the optimal action. 
As a result, existing metrics fall short in preserving decision consistency under critical operating conditions, where small state discrepancies can induce abrupt decision reversals and compromise the stability of distributed resource orchestration.

\subsection{Semantic-Empowered Collaborative Offloading}
While early works on computation offloading relied on Lyapunov optimization or conventional Deep Reinforcement Learning (DRL)~\cite{9449944, zhou2021deep,qi2025efficientinformationupdatescomputefirst}, recent trends focus on joint communication-computation optimization empowered by semantics. Approaches include prioritizing semantically rich task data segments~\cite{zhang2023drl}, dynamically compressing task payloads for bandwidth efficiency~\cite{10508293}, and balancing latency via semantic feature extraction~\cite{chen2025onlinemultitaskoffloadingsemanticaware}.
\textit{Nevertheless}, these methods primarily optimize the task data flow (payload) rather than the control signaling flow (system state). Furthermore, pure data-driven policies (e.g., MAPPO) often suffer from the ``black-box'' nature of neural networks: They lack explicit physical constraints and often exhibit performance oscillation (Ping-Pong effect) near system capacity limits. This limitation underscores the need for collaboration mechanisms that explicitly account for decision consistency under partial observability and system constraints, ensuring stable offloading behavior even when state information is stale or imperfect.

\section{System Model}
\label{sec:system_model}
\begin{figure}[htbp]
    \centering
    \includegraphics[width=0.85\linewidth]{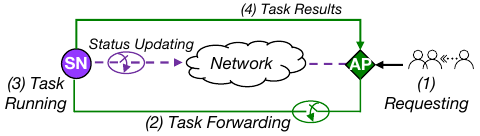}
    \caption{The atomic end-edge collaborative unit in CFN.}
    \label{fig:system-model}
\end{figure}

In this work, we focus on the fundamental ``Atomic Collaboration Unit'' of CFN \cite{RN2762,RN2650}, comprising a single Service Node (SN) and an Access Point (AP). The point-to-point (P2P) edge-end link constitutes the fundamental topological unit wherein the trade-off between status freshness and bandwidth utilization is most pronounced. While real-world CFN architectures may involve multi-hop mesh topologies, the semantic negotiation and decision principles established within this atomic unit serve as the modular foundation for scalable network-wide orchestration.
As illustrated in Fig. \ref{fig:system-model}, the SN and AP collaborate via a closed-loop mechanism comprising two parallel and coupled processes:
\begin{itemize}
    \item \textit{Information Flow (Status Updating):} The SN monitors physical resource states and selectively transmits updates to refresh the status cache at the AP.
    \item \textit{Task Flow (Offloading Decision):} Upon task arrival, the AP utilizes the locally cached state to determine whether to execute the task locally or offload it to the SN.
\end{itemize}

These two flows interact continuously to optimize computing resource utilization and decision timeliness.
\subsection{Network and Computation Model}
The AP and SN communicate through a network connection, which can be characterized as a link exhibiting properties that vary over time. Time is discretized into discrete intervals or slots, denoted as $t\in \{1,2,\dots,T\}$.

\subsubsection{Task Arrival Process}
The task flow arrives at the AP. The $k$th task is represented as a tuple $\Phi_k = \langle t_k, d_k, c_k, \tau_k \rangle$, where:
\begin{itemize}
    \item $t_k$: Task arrival time. To simulate traffic in real network environments, we assume the task inter-arrival time $A_k = t_k - t_{k-1}$ follows a general stochastic distribution $\mathcal{D}_{arr}$.
    \item $d_k$: Input data size (bits).
    \item $c_k$: Required number of computation cycles (cycles).
    \item $\tau_k$: Maximum tolerable latency (Deadline). If the task completion time $t^{(k)}_{comp} > t_k + \tau_k$, the task is considered failed.
\end{itemize}
\subsubsection{Dual-side Computation}
\begin{itemize}
    \item \textit{AP (End-side):} Equipped with limited local computing capabilities (frequency $f_{ap}$, number of cores $N_{ap}$). The AP employs a First-Come-First-Served (FCFS) strategy for its local waiting queue, denoted as $\mathcal{Q}_{ap}$, whose real-time size is represented by $Q_{ap}$ and whose capacity limit is $Q_{ap}^{max}$. For task $\Phi_k$, if local execution is decided ($a_{k}^{ap}=1$), its theoretical processing latency is given by $T_{proc,loc}^{(k)} = c_k / f_{ap} + T_{queue}^{\text{ap}}(t)$, where $T_{queue}^{\text{ap}}(t)$ denotes the residual processing time of the AP queue upon the arrival of task $k$.
    \item \textit{SN (Edge-side):} Equipped with strong computing capabilities (frequency $f_{sn}$, number of cores $N_{sn}$). The SN maintains a task queue $\mathcal{Q}_{sn}$ using the FCFS policy, with a maximum capacity of $Q_{sn}^{max}$ and an real-time size $Q_{sn}$. Similarly, the processing time for task $\Phi_k$ is represented as $T_{proc,sn}^{(k)}$. Tasks will be dropped if the queue overflows or execution times out.
\end{itemize}
\subsubsection{Logical Communication Links}
The connection between the AP and the SN is represented by two logical links that exhibit time-varying properties:
\begin{itemize}
    \item \textit{Uplink (Status updating):} The SN sends resource status updates to the AP. The transmission latency $D_{up}(t)$ is influenced by the time-varying uplink available bandwidth $B_{up}(t)$ and the status packet size $S_{stat}$:
    \begin{equation}
        D_{up}(t) = \frac{S_{stat}}{B_{up}(t)} + D_{prop}(t),
    \end{equation}
    where $D_{prop}(t)$ represents the sum of the time-varying network propagation delay and the processing delay at intermediate routing nodes.
    \item \textit{Downlink (Task offloading):} The AP transmits task data to the SN. The transmission latency is determined by the task data size $d_k$ and the time-varying downlink available bandwidth $B_{down}(t_k)$:
    \begin{equation}\label{eq:d_down}
       D_{down}^{(k)} = \frac{d_k}{B_{down}(t_k)} + D_{prop}(t_k).
    \end{equation}
\end{itemize}
\subsection{Raw Resource State Space}\label{sect:raw-res-state-space}
At time $t$, the physical resource state of the SN is described by a $d_{in}$-dimensional vector $\mathbf{x}_t = [x_{t,1}, x_{t,2}, \dots, x_{t,d_{in}}]^\top \in \mathbb{R}^{d_{in}}$. To comprehensively capture the transient characteristics and long-term trends of the edge server under heterogeneous workloads, we categorize the state features into three dimensions: Computational availability, congestion backlog, and traffic evolution.

\subsubsection{Computational Availability} 
Reflects the SN's ability to immediately process newly arriving tasks at the current moment.
\begin{itemize}
    \item $x_{t,1} = \tilde{n}_{idle}$: Number of Idle Cores. Defined as the number of currently idle CPU cores. This metric directly reflects whether a task can be immediately processed in parallel without entering the queue.
\end{itemize}

\subsubsection{Congestion Backlog} 
Reflects the current load of the system, used to infer queuing delay.
\begin{itemize}
    \item $x_{t,2} = \tilde{q}_{len}$: Normalized Queue Length. Defined as $\frac{L_{queue}(t)}{N_{sn}}$, i.e., the average number of waiting tasks allocated per computing core. It measures the ``depth'' of congestion.
    \item $x_{t,3} = w_{head}$: Head-of-Line (HOL) Waiting Time. The duration the first task in the queue has already waited. In contrast to queue length, this metric more accurately represents the stagnation in current processing speeds and provides an effective means of identifying significant congestion due to HOL blocking, such as when large tasks obstruct the queue.
    \item $x_{t,4} = \delta_{AoI}$: AoI. The time elapsed since the status received by the AP was generated (accurately recorded via the ACK mechanism). Specifically, it represents the duration since the status was last updated from the SN's own perspective, assisting in judging status timeliness.
\end{itemize}

\subsubsection{Traffic Evolution} 
Reflects load variations, used to predict future congestion trends.
\begin{itemize}
    \item $x_{t,5} = \tilde{c}_{last}$: Last Task Workload. The normalized computational volume of the last completed task. This metric captures task flow diversity, indicating whether the system is currently processing intensive or lightweight tasks.
    \item $x_{t,6} = \tilde{A}_{est}$: Estimated Arrival Interval. The moving average of task inter-arrival times. This metric serves as a traffic intensity indicator: A sudden drop in $\tilde{A}_{est}$ implies a surge in task arrival rate, signaling the potential onset of congestion.
\end{itemize}

Nevertheless, a direct dependence on these raw observations results in a semantic gap. The physical measurements are affected by stochastic high-frequency fluctuations, such as transient oscillations in $x_{t,1}$, and exhibit intricate nonlinear relationships, such as the interaction between backlog $x_{t,2}$ and arrival trends $x_{t,6}$. These elements hinder the accurate discernment of the inherent service capability essential for effective offloading decisions.

\subsection{Problem Definition}
The core objective of the CFN system is to mitigate system risk caused by decision-making under partial and stale state information. In particular, inconsistent or outdated status views at the AP may lead to unsafe offloading decisions, resulting in task failures, excessive latency, or unnecessary consumption of communication resources. To address this challenge, we jointly optimize the status update policy $\pi_{sn}$ at the SN and the task offloading policy $\pi_{ap}$ at the AP, such that decision consistency is preserved while operating under communication and computation constraints.

Formally, we characterize system risk through a long-term expected objective that captures the cumulative impact of decision errors, communication overhead, and state inconsistency. The joint optimization problem is formulated as
\begin{equation}
\label{eq:opt-target}
\begin{split}
    \min_{\pi_{sn}, \pi_{ap}} J = & \lim_{T \to \infty} \frac{1}{T} \sum_{t=1}^{T} \bigg[ \mathcal{C}_{task}(t) \\
    & + \lambda_{comm} \mathcal{C}_{comm}(t) + \lambda_{sem} \mathcal{C}_{sem}(t) \bigg],
\end{split}
\end{equation}
where $\lambda_{comm}$ and $\lambda_{sem}$ are non-negative weighting coefficients for communication cost and semantic consistency cost, respectively, used to regulate the trade-off between different optimization objectives. The definitions of each cost term are as follows:

\begin{enumerate}
    \item Task Utility Cost $\mathcal{C}_{task}(t)$:
    The task utility cost $C_{\text{task}}(t)$ captures the direct system-level consequences of unsafe or inconsistent offloading decisions. Task failures and deadline violations reflect instances where the AP’s decision, made under incomplete or outdated state information, is incompatible with the actual execution capability of the system.
    \begin{equation}
    \begin{split}
        \mathcal{C}_{task}(t) = \sum_{k \in \mathcal{K}_t} \Big[ & \mathbb{I}_{fail}^{(k)} P_{fail} \\
        & + (1-\mathbb{I}_{fail}^{(k)}) \cdot \max(0, T_{delay}^{(k)} - \tau_k) \Big],
    \end{split}
\end{equation}
    where,
    \begin{itemize}
        \item $\mathcal{K}_t$ is the set of tasks completed or failed within time slot $t$.
        \item $\mathbb{I}_{fail}^{(k)}$ is an indicator function that takes the value 1 when task $k$ fails due to queue overflow or timeout, and 0 otherwise.
        \item $P_{fail}$ is a constant representing the fixed penalty for task failure.
        \item $T_{delay}^{(k)}$ is the actual end-to-end latency of task $k$.
    \end{itemize}
    \item Communication Cost $\mathcal{C}_{comm}(t)$:
    The communication cost $C_{\text{comm}}(t)$ models the risk of excessive status synchronization and task transmission. Frequent updates and offloading actions consume limited network resources and may interfere with service traffic, indirectly increasing the likelihood of future decision errors due to congestion and delayed feedback.
    \begin{equation}
        \mathcal{C}_{comm}(t) = a_t^{sn} \cdot \omega_{up} + \sum_{k \in \mathcal{K}_t} (1-a_k^{ap}) \cdot \omega_{down},
    \end{equation}
    where, 
    \begin{itemize}
        \item $a_t^{sn} \in \{0, 1\}$ is the update action of the SN in time slot $t$ (1 indicates sending an update).
        \item $a_k^{ap} \in \{0, 1\}$ is the decision action of the AP for task $k$ (0 indicates offloading, 1 indicates local execution).
        \item $\omega_{up}$ is the transmission penalty coefficient for status update, used to quantify the bandwidth resource cost occupied by transmitting a semantic state packet. A larger $\omega_{up}$ implies uplink congestion or bandwidth scarcity, forcing the system to allow updates only when the semantic value is extremely high.
        \item $\omega_{down}$ is the transmission penalty coefficient for task offloading, representing the bandwidth resource cost required to transmit a task over the downlink. This coefficient maps ``network resource consumption'' to the same cost dimension as ``latency,'' achieving a unified quantification of communication overhead and computation latency under the same objective function.
    \end{itemize}

    \item Semantic Consistency Cost $\mathcal{C}_{sem}(t)$:
    The semantic consistency cost $C_{\text{sem}}(t)$ regularizes the temporal evolution of abstracted system states and penalizes abrupt representation changes induced by noise. From a systemic viewpoint, this term functions to limit instability within the decision-making foundation, thereby mitigating the likelihood of oscillatory or conflicting offloading choices that may result from rapid fluctuations in system states.
\end{enumerate}

Together, these cost components provide a unified risk-aware formulation that links state synchronization, decision-making, and system performance. Minimizing the objective in (\eqref{eq:opt-target}) promotes decision consistency over time, particularly in high-load regimes where small state discrepancies can induce disproportionate system-level failures.

\section{SenseCFN}
\label{sec:methodology}
\begin{figure}[htbp]
    \centering
    \includegraphics[width=\linewidth]{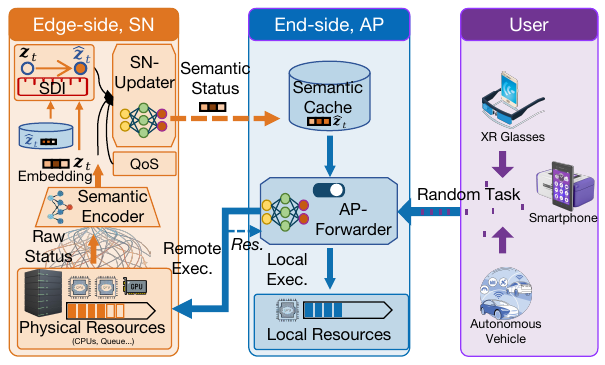}
    \caption{SenseCFN Framework.}
    \label{fig:arch}
\end{figure}

As shown in Fig. \ref{fig:arch}, the SenseCFN framework comprises three core models: (1) The \textit{Semantic Encoder} utilized at the SN is engineered to encode complex raw statuses into semantic embeddings. (2) The \textit{Collaborative Dual-Decision Network}, which consists of the SN-Updater and the AP-Forwarder, autonomously executes status updates and task forwarding decisions.
Subsequently, we employ the CTDE algorithm, which leverages global information to enhance the performance of the previously mentioned models in an offline setting.

\subsection{Semantic Encoder}
\label{subsec:semantic_space}
To mask the diversity of underlying heterogeneous resources and effectively capture the time-varying nature of workloads, the primary task of SenseCFN is to construct a semantic state manifold that bridges the physical resource layer and the upper decision layer. It is crucial to note that the framework proposed is not restricted by the choice of encoder architecture; although the Transformer is utilized to effectively capture nonlinear state coupling, it is not an essential component.

\subsubsection{Transformer-based Temporal Status Encoding}\label{sect:encoder}
To extract latent features useful for AP decision-making from $\mathbf{x}_t$ and to capture the nonlinear characteristics and dynamic evolution patterns of resource states, we design a Transformer-based encoder, denoted as $E_{\theta}$, as the backbone network for feature extraction.

The input to $E_{\theta}$ is the historical state sequence $\mathbf{X}_t \in \mathbb{R}^{w \times d_{in}}$ within a time window. The data stream first passes through a Linear Projection layer to be mapped into a high-dimensional embedding space and is injected with Positional Encoding to preserve sequential order information. Subsequently, the data passes through $N$ stacked Transformer Encoder Blocks. Each block consists of two sub-layers: Multi-Head Self-Attention (MSA) and a Feed-Forward Network (FFN), supplemented by Layer Normalization (LayerNorm) and Residual Connections:
\begin{align}
    \mathbf{H}'_l &= \text{MSA}(\text{LayerNorm}(\mathbf{H}_{l-1})) + \mathbf{H}_{l-1}, \\
    \mathbf{H}_l &= \text{FFN}(\text{LayerNorm}(\mathbf{H}'_l)) + \mathbf{H}'_l.
\end{align}
Finally, a global pooling layer aggregates the sequence features into a $d_{sem}$-dimensional semantic encoding $\mathbf{z}_t = E_{\theta}(\mathbf{x}_t) \in \mathbb{R}^{d_{sem}}$. 

In this paper, we employ the Transformer encoder to map the state to a 3-dimensional semantic representation, compressing the state dimensions while preserving its semantics. The encoder consists of $L=2$ encoding layers with $H=4$ attention heads per layer, a model width of $64$, and a feed-forward layer width of $256$, with a dropout rate of $p = 0.1$.

\textit{Discussion on Observation Window and Temporal Dependency}:
$E_{\theta}(\cdot)$ is designed as a general-purpose sequence processing model that naturally supports inputs of arbitrary length $w > 1$, capable of capturing long-range dependencies via the MSA mechanism. In our experimental configuration, we set $w=1$ to validate the model's instant inference capability under minimal communication overhead.
The validity of this setting relies on the following fact: The input features defined in Section \ref{sect:raw-res-state-space} (e.g., ``Queue Length'' and ``HOL Waiting Time'') fall under the category of cumulative statistics in queuing theory. They are inherently the integral results of the difference between historical arrival and service rates, implicitly encapsulating the system's historical state information. Consequently, even under this Markov approximation setting ($w=1$), the powerful nonlinear fitting capability of the Transformer (specifically the FFN) allows it to effectively decouple the workload evolution trend from the current snapshot, without requiring an explicit long-sequence input.

\subsubsection{Semantic Deviation Index (SDI)}\label{sect:sdi}
To realize event-driven communication, it is necessary to define a metric in the latent space to quantify the ``semantic distance'' between the current state and the state perceived by the receiver. Based on the feature encoding $\mathbf{z}_t$, we define the SDI as follows:
\begin{equation}
    \text{SDI}_t = \frac{\| \mathbf{z}_t - \hat{\mathbf{z}}_t \|_2}{\| \hat{\mathbf{z}}_t \|_2 + \epsilon},
\end{equation}
where $\hat{\mathbf{z}}_t$ is the semantic encoding currently cached at the AP (which can be obtained in real-time by the SN from the transmission history), and $\epsilon$ is a small value to prevent division by zero. 

The SDI essentially calculates the normalized relative error between two vectors in Euclidean space. It acts as a trigger: The SDI surges significantly to activate a status update only when a state change induces a substantial displacement in the semantic space (i.e., exerting a substantive impact on decision-making).

\subsubsection{Semantic Manifold Smoothness}\label{sect:semantic-smooth}
Since raw states $\mathbf{x}_t$ (especially the idle core count $x_{t,1}$) often contain high-frequency jitter noise, direct mapping may cause severe oscillation of the semantic vector $\mathbf{z}_t$ within the feature space. 

To augment the robustness of semantic representation, a manifold smoothness constraint is implemented in the feature extraction phase. This constraint enforces a time continuity prior, ensuring that the state manifold developed by the encoder undergoes smooth evolution. The mathematical expression is specified as follows:
\begin{equation}
    \label{eq:sem_reg}
    \mathcal{R}_{smooth}(\mathbf{z}_t, \mathbf{z}_{t-1}) = \| \mathbf{z}_t - \mathbf{z}_{t-1} \|_2^2.
\end{equation}

\subsection{Collaborative Dual-Decision Network}
\label{subsec:dual_policy}
In order to facilitate collaboration between end and edge devices amidst limited communication capacities, we introduce two lightweight neural networks. These networks are implemented at the SN and AP, respectively. Despite their asynchronous physical operation, they attain implicit collaboration by utilizing a common semantic state space.

\subsubsection{SN-Updater}
The core task of the SN-Updater is to determine, at each time slot $t$, whether to consume bandwidth to transmit the latest semantic vector. To reduce inference energy consumption at the edge and accelerate convergence, we design a lightweight Multilayer Perceptron (MLP) classifier $\pi_{sn}(\cdot; \theta_{sn})$ and train it using a supervised imitation learning scheme.

\begin{itemize}
    \item Input Features: The network input is a composite feature vector:
        \begin{equation}
            \label{eq:sn_input}
            \mathbf{s}_{sn, t} = [\mathbf{z}_t \oplus \text{SDI}_t \oplus \text{QoS}_t],
        \end{equation}
        where $\oplus$ denotes vector concatenation. $\mathbf{z}_t$ provides the contextual semantics of the current workload; $\text{SDI}_t$ serves as an explicit ``update urgency'' feature, enabling the network to perceive state deviations without relearning complex comparison logic. Notably, $\text{QoS}_t$ is an uplink congestion indicator based on local observations. By monitoring the SN's local buffer backlog and historical bandwidth estimates, it quantifies the communication cost (transmission delay estimation) of triggering an update. Introducing this feature endows the policy network with cost-awareness, automatically suppressing non-urgent updates during network congestion.
    
    \item Network Output: The output layer of $\pi_{sn}$ passes through a Sigmoid activation function to generate a scalar probability $p_{up} \in [0, 1]$. This value quantifies the confidence in the necessity of triggering a semantic update at the current moment.
    
    \item Trigger Decision: The final discrete action $a_t^{sn} \in \{0, 1\}$ is determined by the network output probability:
    \begin{equation}
        \label{eq:sn_decision}
        a_t^{sn} = 
        \begin{cases} 
        1 \ (\text{Update}), & \text{if } p_{up} > 0.5 \\ 
        0 \ (\text{Wait}), & \text{otherwise} 
        \end{cases}.
    \end{equation}
\end{itemize}

\subsubsection{AP-Forwarder}
The task of the AP agent is to select the optimal execution location $a_k^{ap} \in \{0, 1\}$ for the arriving task $k$, where 1 represents local execution and 0 represents offloading to the SN. Considering the computational resource constraints of end-side devices and the potential staleness of the semantic state $\hat{\mathbf{z}}$, we design the network $\pi_{ap}(\cdot; \theta_{ap})$ as a lightweight MLP classifier with information timeliness awareness. This architecture aims to efficiently process heterogeneous input information and achieve robust decision-making while ensuring real-time inference.

\begin{itemize}
    \item Input Features: To comprehensively perceive the decision environment, the input vector $\mathbf{s}_{ap, k}$ directly fuses high-dimensional semantic features with low-dimensional physical states:
    \begin{equation}
        \mathbf{s}_{ap, k} = [\hat{\mathbf{z}}_{t_k} \oplus \delta_{AoI, t_k} \oplus \hat{D}_{down}(t_k) \oplus \mathbf{x}_{loc}(t_k)],
    \end{equation}
    where $\oplus$ denotes the vector concatenation operation. $\hat{\mathbf{z}}_{t_k}$ is the cached semantic vector, $\delta_{AoI, t_k}$ is its normalized AoI, and $\mathbf{x}_{loc}(t_k)$ contains the AP's local resource state (e.g., queue length and idle core count).
    
    \item Network Output and Decision: The network outputs a scalar $p_{loc} \in [0, 1]$, representing the confidence in retaining the task locally. The final offloading action $a_k^{ap}$ is generated as follows:
    \begin{equation}
        \label{eq:ap_decision}
        a_k^{ap} = 
        \begin{cases} 
        1 \ (\text{Local}), & \text{if } p_{loc} > 0.5 \\ 
        0 \ (\text{Offload}), & \text{otherwise} 
        \end{cases}.
    \end{equation}
\end{itemize}

Departing from hard-coded gating structures, SenseCFN adopts a joint state representation scheme. Although the input features are simple direct concatenations, driven by end-to-end training, the MLP network can adaptively learn nonlinear interaction relationships between features via gradient updates. Specifically, the network can implicitly establish a negative correlation mapping between $\delta_{AoI, t_k}$ and the credibility of $\hat{\mathbf{z}}_{t_k}$: When $\delta_{AoI}$ is large, the neurons inside the network automatically suppress the activation weights of semantic features, shifting reliance towards local states $\mathbf{x}_{loc}$ for inference. This data-driven soft attention characteristic allows the AP to achieve robust performance and smooth degradation of decision-making in the absence of explicit filtering modules, even when information is stale.

\subsection{Semantic CTDE algorithm}
\label{subsec:ctde}
To efficiently acquire the collaborative dual-decision network, we propose a Semantic CTDE algorithm. Leveraging the global system state available during the offline training phase, this algorithm transforms the policy optimization problem into a fully end-to-end supervised learning and system optimization problem. 

\subsubsection{Training Architecture and Data Flow}
In the centralized training process, we connect the semantic encoder $E_{\theta}$, the SN-Updater $\pi_{sn}$, and the AP-Forwarder $\pi_{ap}$ into a differentiable computational graph. The data flow is as follows:
\begin{enumerate}
    \item \textit{Semantic Extraction}: Encoder $E_{\theta}$ receives the raw resource state $\mathbf{x}_t$ and outputs the semantic vector $\mathbf{z}_t = E_{\theta}(\mathbf{x}_t)$.
    \item \textit{SN Decision}: The signal $\mathbf{s}_{sn,t}$, incorporating $\mathbf{x}_t$, is processed by $\pi_{sn}$ to generate the update decision $a^{sn}_t$.
    \item \textit{State Evolution Simulation}: The trainer simulates the communication process based on $a^{sn}_t$. If an update occurs ($a^{sn}_t = 1$), the AP's cache $\hat{\mathbf{z}}$ is refreshed to $\mathbf{z}_t$; otherwise, it retains the value from the previous moment. 
    \item \textit{AP Decision}: $\pi_{ap}$ receives the updated joint state $\mathbf{s}_{ap,k}$ and outputs the forwarding decision $a^{ap}_k$.
\end{enumerate}

\subsubsection{Hybrid Label Generation with Physical Bound Calibration}
\label{sec:label_generation}
To generate effective supervision signals, we employ a ``dual-decision filtering'' approach that combines offline expert execution traces for policy initialization with physical posterior constraints for relabeling.

\paragraph{AP Forwarding Decision}
We construct the offloading label $y_{ap, k}^*$ (1 for local, 0 for offloading) via a two-layer calibration process. For data derived from mature baselines, we directly employ behavioral cloning to establish a solid performance foundation. For exploration data, we enforce safety boundaries where mandatory local execution ($y_{ap, k}^*=1$) is assigned if specific fail-safe conditions are met. Specifically, a congestion circuit breaker ($Q_{sn} \ge \delta_{cong}$) is used to physically isolate tasks from edge saturation to prevent timeouts. Furthermore, we apply hysteresis control ($(C_{rem} - C_{loc}) \ge \epsilon_{hyst}$) to suppress Ping-Pong effects caused by bandwidth jitter, ensuring decisions only reverse when the remote cost penalty is structurally significant.

\paragraph{SN Update Decision}
The SN update label $y_{sn, t}^*$ is modeled as an event-driven process balancing timeliness and overhead. An update is triggered ($y_{sn, t}^* = 1$) if either the maximum AoI constraint ($\text{AoI}_t \ge \tau_{max}$) is breached to prevent cache staleness, or a critical congestion warning ($Q_{sn} \ge \delta_{warn}$) indicates a high-value ``semantic shift.'' This mechanism ensures the AP perceives sudden edge congestion timely while minimizing unnecessary transmissions during stable periods.

\subsubsection{Joint Differentiable Objective Function}
The original optimization objective (Eq. \eqref{eq:opt-target}) contains discrete action variables and hard constraints, making it non-differentiable and difficult to use directly for neural network backpropagation. We transform the system's multi-dimensional optimization objective into the following joint loss function $\mathcal{L}_{total}$, balancing imitation learning and system cost optimization via weighted summation:

\begin{equation}
    \label{eq:total_loss}
    \mathcal{L}_{total} = \mathcal{L}_{imit} + \mathcal{L}_{sys} + \lambda_{sem} \mathcal{L}_{cons}.
\end{equation}

\paragraph{Imitation Learning Loss $\mathcal{L}_{imit}$}
This term fits the aforementioned expert labels, adopting the Cross-Entropy form:
\begin{equation}
    \mathcal{L}_{imit} = \lambda_{r} \cdot \text{CE}(\pi_{sn}, y_{sn}^{*}) + \lambda_{ap} \cdot \text{BCE}(\pi_{ap}, y_{ap}^{*}),
    \label{eq:limit_loss}
\end{equation}
where $\text{CE}(y,p)=-\sum_{c=1}^M{y_c \log (p_c)}$ is the cross entropy loss, $\text{BCE}(y,p)=-(y\cdot \log(p) + (1-y) \cdot \log (1-p))$ is the binary cross entropy loss, and $\lambda_{r}$ and $\lambda_{ap}$ are weighting coefficients. This ensures the policy rapidly acquires basic usability during the cold start phase.

\paragraph{System Cost Loss $\mathcal{L}_{sys}$}
To break through the limitations of expert rules and directly optimize end-to-end performance, we introduce physical cost constraints containing success rate inference:
\begin{equation}
    \mathcal{L}_{sys} = \lambda_{inf}\mathcal{L}_{succ} + \lambda_{c}\overline{p}_{up} + \lambda_{f}\mathcal{C}_{forward} + \lambda_{lat}\mathcal{L}_{delay}.
    \label{eq:sys_loss}
\end{equation}
The specific definitions are as follows:
\begin{itemize}
    \item \textit{Success Rate Inference Regularization ($\mathcal{L}_{succ}$)}: To explicitly maximize the task success rate, we introduce this term as an auxiliary supervision signal. It measures the cross-entropy between the AP decision probability distribution and the final task success status (Success/Fail):
    \begin{equation}
        \mathcal{L}_{succ} = \mathbb{E}_{\mathcal{B}} [ \text{BCE}(p_{loc}, y_{success}) ],
    \end{equation}
    where $y_{success} \in \{0, 1\}$ indicates whether the task is ultimately completed successfully. Minimizing this term guides the model to converge its probability distribution towards the high success rate decision space.

    \item \textit{Average Communication Overhead ($\bar{p}_{up}$)}: Determined by the mean of the update probability $p_{up} = \pi_{sn}(\mathbf{s}_{sn})_0$ output by the SN network over the training batch, i.e., $\bar{p}_{up} = \mathbb{E}_{\mathcal{B}}[p_{up}]$. Minimizing this term suppresses unnecessary communication frequency.
    
    \item \textit{Expected Execution Latency ($\mathcal{C}_{forward}$)}: Leveraging the local execution probability $p_{\text{loc}}$ output by the AP network, we weight the physical time consumption of local and remote execution:
    \begin{equation}
        \mathcal{C}_{forward} = \mathbb{E}_{\mathcal{B}} \left[ p_{\text{loc}} \cdot T_{\text{loc}} + (1 - p_{\text{loc}}) \cdot T_{\text{off}} \right],
    \end{equation}
    where $T_{\text{loc}}$ is the normalized actual execution time locally, and $T_{\text{off}}$ is the total time for offloading to the SN (including transmission and queuing). This term establishes a differentiable link between decision probability and physical latency, directing gradient descent towards lower latency.
    
    \item \textit{Latency Penalty ($\mathcal{L}_{delay}$)}: Applies a soft penalty to tasks exceeding the deadline $\tau$:
    \begin{equation}
        \mathcal{L}_{delay} = \mathbb{E}_{\mathcal{B}} \left[ \frac{\text{ReLU}(T_{actual} - \tau)}{\tau + \epsilon} \right].
    \end{equation}
    This term forces the model to treat satisfying timeliness constraints as a hard premise when trading off energy consumption and latency.
\end{itemize}

\paragraph{Semantic Consistency Loss ($\mathcal{L}_{cons}$)}
This term directly references the manifold smoothness regularization term \eqref{eq:sem_reg} defined earlier, i.e., $\mathcal{L}_{cons} = \mathcal{R}_{smooth}$. It serves as a temporal stability constraint for the semantic encoder, forcing the latent space manifold to remain locally smooth. This ensures that displacements in semantic vectors are primarily driven by structural changes in system load (rather than random noise), thereby providing a robust feature basis for the convergence of collaborative policies.

\begin{algorithm}[t]
\caption{Semantic CTDE Training Process}
\label{alg:semantic_ctde}
\SetAlgoLined
\DontPrintSemicolon
\KwIn{Raw traces $\mathcal{T}_{raw}$, thresholds $\delta_{cong}, \delta_{warn}, \epsilon_{hyst}, \tau_{max}$}
\KwOut{Trained parameters $\theta$ for $E_\theta, \pi_{sn}, \pi_{ap}$}

\tcp{Phase 1: Offline Hybrid Label Generation}
Initialize $\mathcal{D} \gets \emptyset$\;
\For{each time step $t$ in $\mathcal{T}_{raw}$}{
    $y_{sn, t}^* \gets \mathbb{I}(\text{AoI}_t \ge \tau_{max} \lor \mathcal{Q}_{sn}^{(t)} \ge \delta_{warn})$\;
    \lIf{Expert $a_t$ exists}{$y_{ap, t}^* \gets a_t$}
    \lElseIf{$\mathcal{Q}_{sn}^{(t)} \ge \delta_{cong} \lor \Delta C_{cost} \ge \epsilon_{hyst}$}{$y_{ap, t}^* \gets 1$}
    \lElse{$y_{ap, t}^* \gets 0$}
    $\mathcal{D} \gets \mathcal{D} \cup \{(s_t, y_{sn, t}^*, y_{ap, t}^*)\}$\;
}

\tcp{Phase 2: Centralized Training}
Initialize $\theta$ with random weights\;
\For{epoch $= 1$ to $N_{epochs}$}{
    \For{minibatch $\mathcal{B} \subseteq \mathcal{D}$}{
        Generate latents $z_t, \hat z_t$ via $E_\theta$ and compute $SDI_t$\;
        Compute joint loss $J_{total}$ (Eq. \ref{eq:total_loss})\;
        Update $\theta \gets \text{Adam}(\theta, \nabla_\theta J_{total})$\;
    }
}
\end{algorithm}

The training procedure, summarized in Algorithm \ref{alg:semantic_ctde}, proceeds in two phases. 
In Phase 1 (Lines 1--8), we construct a labeled dataset $\mathcal{D}$ via offline calibration using physically grounded thresholds. The SN update label $y_{sn}^*$ is generated using an indicator function based on a timeliness limit of $\tau_{max}=0.5$s (approx. 1/4 of the end-to-end deadline) and a congestion warning threshold of $\delta_{warn}=0.5$ (indicating 50\% resource occupancy) (Line 3). For the AP offloading label $y_{ap}^*$, we prioritize expert traces if available (Line 4); otherwise, we apply physical safety bounds (Line 5), assigning mandatory local execution ($y_{ap}^*=1$) when the circuit breaker triggers at $\delta_{cong}=0.5$ (reserving a 50\% safety buffer) or when the offloading advantage $\Delta C_{cost}$ falls below the hysteresis margin $\epsilon_{hyst}=0.05$s (approx. $1.5\times$ the average task service time), effectively filtering out transient jitter.
In Phase 2 (Lines 9--14), we perform centralized training to optimize the network parameters $\theta$. The models are trained in an end-to-end manner utilizing the Adam optimizer, employing a learning rate of $\eta=10^{-3}$ along with a weight decay of $10^{-4}$, which facilitate stable convergence. We employ a batch size of $|\mathcal{B}|=256$ and train for $N_{epochs}=60$ epochs to allow sufficient fitting of the expert strategies. In each iteration, the joint objective $J_{total}$ (Eq. \eqref{eq:total_loss}) aggregates imitation losses, semantic regularization, and system costs to update $\theta$ (Line 14).

\section{Experiments and Analysis}
\label{sec:experiment}
To comprehensively evaluate the effectiveness of the SenseCFN framework and its semantic collaborative training and inference mechanisms, we developed a Discrete-Event Simulator (DES) based on Python 3.10 and SimPy. 
The experimental environment comprises a computation-constrained end-side AP and a resource-rich SN. Both communication links (including the uplink status flow and downlink task flow) and task generation processes are modeled as stochastic processes. The experiments primarily evaluate the \textit{Task Success Rate} (i.e., tasks completed within the latency threshold) and the \textit{Status Update Frequency} (reflecting the update cost).

\noindent\textbf{Baselines.}
We select three categories of methods for comparison, spanning both temporal and content dimensions and covering traditional rule-based methods as well as reinforcement learning methods:
\begin{enumerate}
    \item \textit{QAoI (PPO)}: We implemented a PPO-based baseline optimizing the Query Age of Information (QAoI) \cite{qaoi2022tcomm}. This metric is currently recognized as the State-of-the-Art (SOTA) mechanism in status update systems, as it focuses on minimizing information staleness specifically at the critical instants of task arrival (Query) rather than average time. Following the original formulation, we limit the token bucket capacity to 50 to constrain update costs while optimizing freshness.
    \item \textit{Content-Aware}: An update policy based on the content difference dimension. Updates are triggered only when the current SN computing resource state (specifically, the number of remaining computing cores) differs from the AP cache. This represents traditional update methods based on rigid constraints of data numerical discrepancy or error \cite{RN1605}.
    \item \textit{Fixed Frequency}: Sends status updates at fixed intervals (50ms, i.e., 20Hz). This is currently the most common benchmark method, lacking capabilities for freshness and content perception.
\end{enumerate}

\noindent\textbf{Dataset.}
Acquiring high-quality training data is one of the core challenges in building task-oriented semantic models. We introduce the generation and labeling of training data.
\subsubsection{Random Task Arrival Data Generation}
To simulate a realistic and challenging edge computing environment, following established conventions in the MEC field \cite{8016573,9829241,RN2650}, we model the task arrival process as a Poisson Process. This stochastic process not only aligns with the statistical characteristics of independent IoT requests, ensuring the theoretical soundness of the experiments, but its core memoryless property also introduces significant stochastic interference. This unpredictability effectively prevents the model from overfitting to simple temporal patterns, forcing the SN and AP to learn update policies based on real-time states amidst uncertainty.

We define the task arrivals as a Poisson process with intensity $\lambda_{in}$, such that the number of arrivals within time $t$ satisfies $N(t)\sim \mathrm{Poisson}(\lambda_{in} t)$. Equivalently, the inter-arrival times $A_k$ follow an exponential distribution, with the mean determined by parameter $\lambda$:
\begin{equation}
  A_k \sim \mathrm{Exponential}(\lambda_{in}), \qquad \mathbb{E}[A_k]=\frac{1}{\lambda_{in}}.
\end{equation}

While Poisson arrivals are used in the experiments, the proposed method makes decisions based on the current system state and does not rely on a specific arrival distribution.

\noindent\textbf{Settings.}
\noindent\textit{Simulation Environment:}
We simulate a heterogeneous edge computing environment consisting of a powerful Server Node (SN) and a resource-constrained Access Point (AP). To reflect real-world hardware disparities, the SN is configured with 4 cores at 1.0 GHz, supporting parallel task execution, while the AP operates with 2 cores at 0.8 GHz. The network is modeled as uplink-constrained, with an uplink bandwidth of $\sim$20 Mbps versus a downlink bandwidth of $\sim$50 Mbps, highlighting the necessity of semantic compression for status updates. Finite queue capacities are enforced (SN: 100 tasks, AP: 25 tasks) to faithfully simulate buffer overflow and packet drop phenomena under congestion.

\noindent\textit{Workload Generation:}
Task arrivals follow a stochastic process with rates $\lambda_{in}$ sweeping from 10 Hz to 60 Hz (step size 10), covering scenarios from light load to severe overload. Each task mimics a computation-intensive AI inference job, where the workload follows a Gaussian distribution ($\mu=10^8$ cycles, $\sigma=2{\times}10^7$) linearly correlated with data size ($\mu=1.0$ MB, $\sigma=0.2$). A hard QoS constraint is imposed with an end-to-end deadline of $1.8$ s; tasks exceeding this limit are deemed failures. Each simulation episode spans $500$ s to ensure statistical stability.

\noindent\textit{Model Implementation:}
The proposed semantic framework is designed for lightweight deployment. The semantic encoder ($E_\theta$) employs a 2-layer Transformer with 4 attention heads, mapping state inputs to a compact latent dimension of 3. Both the SN update policy and AP forwarding policy are implemented as MLP networks ($64{\times}64$ hidden units with GeLU activation). The CTDE framework is trained for 60 epochs with a batch size of 256, using loss weights $\lambda_{ap}=1.2$ and $\lambda_{lat}=0.2$ to prioritize success rate while balancing latency. For baseline comparisons, the PPO update budget is capped at 50 Hz, and the rule-based SDI threshold is set to $\delta_{th}=0.2$.

\noindent\textbf{System Overhead and Deployment Cost.}
To validate the feasibility of deploying SenseCFN on resource-constrained edge devices, we analyze the model complexity. The proposed Transformer-based encoder ($L=2, H=4, d_{model}=64$) contains approximately 1.3$\times$10$^5$ parameters, occupying less than 0.6 MB of memory (FP32). On standard edge processors (e.g., ARM Cortex-A72), the inference latency for such a lightweight model is typically $<$ 2 ms. This computational overhead is negligible (orders of magnitude lower) compared to the network transmission latency and the task execution time (hundreds of ms). 
\subsection{Results}
\label{subsec:results}

\subsubsection{Task Success under Saturation and Update Cost}

\begin{figure}[t]
  \centering
  \begin{subfigure}[t]{0.5\linewidth}
    \centering
    \includegraphics[width=\linewidth]{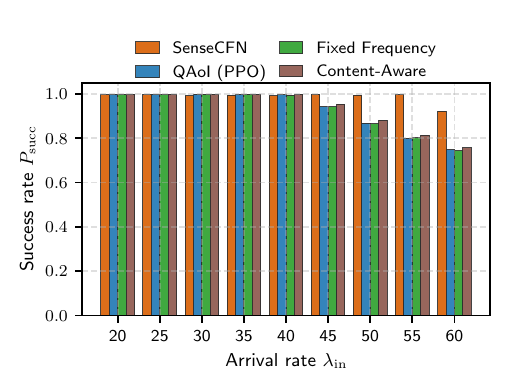}
    \caption{Task success rate.}
    \label{fig:poisson_success_rate}
  \end{subfigure}\hfill
  \begin{subfigure}[t]{0.5\linewidth}
    \centering
    \includegraphics[width=\linewidth]{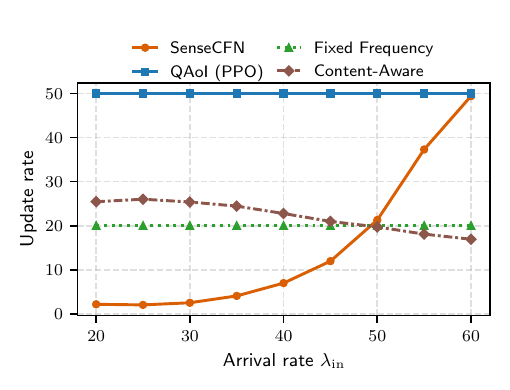}
    \caption{Update frequency.}
    \label{fig:poisson_sn_update_rate}
  \end{subfigure}
  \caption{Impact of different task arrival rates on performance.}
  \label{fig:success_rate}
\end{figure}

As a task-oriented study, we first investigate the task success rate. Fig. \ref{fig:poisson_success_rate} illustrates the variation of the task success rate $P_{succ}$ with respect to the task arrival rate $\lambda_{in}$. The results indicate that in terms of success rate, SenseCFN achieves a success rate approaching 100\% under low or fully load conditions ($\lambda_{in}\leq 55$, approx. 100\% workload), consistent with baseline methods. When the load increases to 55, the system approaches saturation (based on system computing capacity and task size settings, the processing rate at full load is approx. 56 tasks/s). Here, SenseCFN demonstrates a notable improvement in task success rate. Particularly in the high-load scenario where $\lambda_{in} =55$, SenseCFN improves by 25\% ($P_{succ}=99.6\%$) compared to the second best baseline (Content-Aware, 81.3\%), validating the effectiveness of semantic-encoding-driven updates for task computing purposes.

Simultaneously, Fig. \ref{fig:poisson_sn_update_rate} shows the variation of status update frequency with task arrival rate $\lambda_{in}$. Overall, in the lower task arrival range ($\lambda_{in}<40$), SenseCFN's update frequency remains consistently lower than that of baseline methods. Specifically, within the range of $\lambda_{in} \leq 40$, SenseCFN's update frequency peaks at merely 6.97 updates/s (at $\lambda_{in}=40$). At $\lambda_{in}=15$, SenseCFN records only 1.95 updates/s. This represents a reduction of approximately 69.3\% compared to Content-Aware (22.67 updates/s) and roughly  96.1\% compared to QAoI (PPO) (50 updates/s), demonstrating that SenseCFN significantly outperforms the baseline algorithms. 
Combining this with Fig. \ref{fig:poisson_success_rate}, even in the saturation zone of 55 tasks/s, SenseCFN maintains a 99.6\% success rate with an update frequency of 37.3 updates/s. This corresponds to an update-to-task ratio of approximately $37.3/55 \approx 0.678$ under full load. It is evident that SenseCFN maintains a high success rate with a lower update frequency under low loads, while autonomously increasing the update frequency under high loads to enhance success rates, achieving flexible adaptation. Remarkably, SenseCFN's update frequency remains consistently lower than the task arrival rate throughout. This implicitly verifies the prevalence of redundant updates within the system. Furthermore, the baseline methods show little difference in success rates due to their lack of perception regarding task effects, further proving the effectiveness of our proposed method.

It is worth noting that the comparative method QAoI (PPO) constantly hovers at the update frequency threshold of 50 updates/s, which is closely related to the difficulty of learning Poisson processes. Conversely, the update frequency of the Content-Aware method decreases in the SN pressure zone. This is because the SN is consistently in a busy state, which increasingly aligns with the SN status perceived by the AP; consequently, the update frequency drops, and for the AP, local execution becomes the prioritized choice for task computation.

\begin{figure}[t]
  \centering
  \begin{subfigure}[t]{0.5\linewidth}
    \centering
    \includegraphics[width=\linewidth]{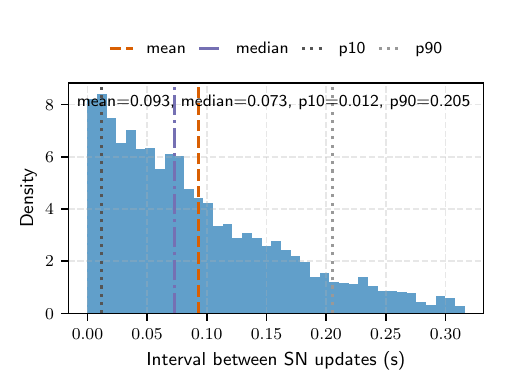}
    \caption{$\lambda_{in}=20$}
  \end{subfigure}\hfill
  \begin{subfigure}[t]{0.5\linewidth}
    \centering
    \includegraphics[width=\linewidth]{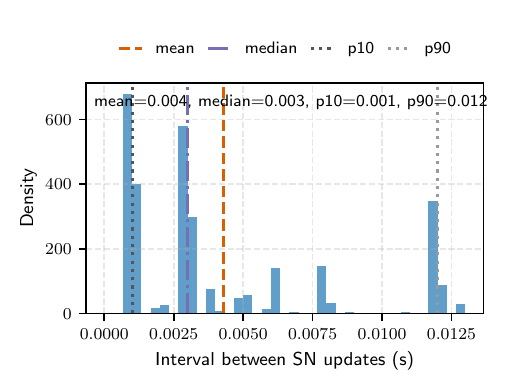}
    \caption{$\lambda_{in}=60$}
  \end{subfigure}
  \caption{SN update interval density distribution (statistics over the full simulation cycle).}
  \label{fig:sn_update_density}
\end{figure}

To verify whether SenseCFN updates occur at random or fixed instances, we statistically analyzed the update intervals over the entire simulation period and plotted the distribution for low load ($\lambda_{in}=20$) and high load ($\lambda_{in}=60$) scenarios, as shown in Fig. \ref{fig:sn_update_density}. It is clearly visible that the update intervals are not fixed, indicating that SenseCFN effectively learns the stochastic patterns of the environmental state and intelligently determines update timing. Meanwhile, as the arrival rate increases, the update interval shortens and the frequency rises, which aligns with expectations.

\subsubsection{Effect of State Abstraction on Offloading Decisions}

\begin{figure}[t]
  \centering
  \begin{subfigure}[t]{0.48\linewidth}
    \centering
    \includegraphics[width=\linewidth]{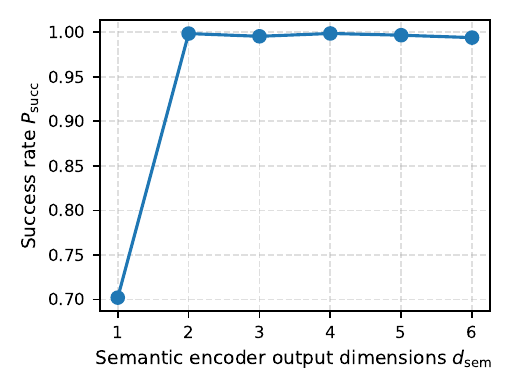}
    \caption{Success Rate}
  \end{subfigure}\hfill
  \begin{subfigure}[t]{0.48\linewidth}
    \centering
    \includegraphics[width=\linewidth]{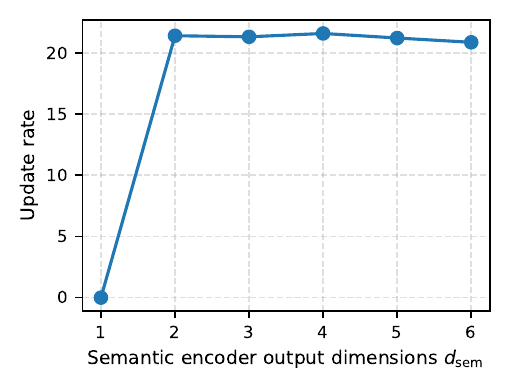}
    \caption{Update Frequency}
  \end{subfigure}
  \caption{Impact of semantic encoder output dimension on success rate under different task arrival distributions.}
  \label{fig:latent_success}
\end{figure}

Converting from the raw resource state $\mathbf{x}_t$ to $\mathbf{z}_t$ offers another advantage besides extracting system state evolution patterns and task arrival laws: it compresses the data volume transmitted between the SN and AP. Therefore, this section tests the dimension (or length) $d_{sem}$ of the transmitted encoding $\mathbf{z}_t$. We increased the encoding dimension from 1 to 6 (the raw state contains 6 dimensions) and calculated the mean success rate at $\lambda_{in}=50$. The results are shown in Fig. \ref{fig:latent_success}.

It can be seen that SenseCFN's success rate is lowest when the encoding dimension is 1, at only 0.7. However, when $d_{sem}$ increases to 2, the success rate recovers to nearly 1, indicating that the encoder's expressive capacity is sufficient to cover semantic features. Considering stability, this paper selects $d_{sem}=3$ as the default value.

\section{Conclusion}
In this paper, we addressed the inefficiency of rigid status updating in task-oriented end-edge collaborative CFN by proposing SenseCFN, a semantic cognitive framework. Unlike traditional approaches that rely on temporal freshness, our framework constructs a semantic representation space mapping directly to decision risks. By leveraging a Transformer-based temporal encoder and the SDI, we established a mechanism where updates are triggered solely by structural semantic shifts that substantively impact downstream decisions. Moreover, a CTDE approach was utilized alongside hybrid label calibration to simultaneously optimize the SN update mechanism and AP offloading strategies while adhering to physical limitations. Simulations demonstrate that SenseCFN significantly reduces communication overhead while enhancing task success rates and system robustness compared to baselines, particularly under high-load stochastic traffic conditions. Future work will extend this semantic paradigm to explore decentralized collaboration among heterogeneous edge clusters.

\balance
\bibliographystyle{unsrt}
\bibliography{refs.bib}

\end{document}